\documentstyle[12pt,epsfig]{article}
\newcommand\mysection{\setcounter{equation}{0}\section}

\def\bm#1{{\mbox{\boldmath $#1$}}}
\def\MSbar{{$\overline{\mbox{\rm MS}}$}}

\def\half{{\textstyle {1\over2}}}

\def\VEV#1{\left\langle #1 \right\rangle}
\def\beq{\begin{equation}}   \def\eeq{\end{equation}}
\def\as{\relax\ifmmode\alpha_s\else{$\alpha_s${ }}\fi}
\def \pt{\relax\ifmmode{p_t}\else{$p_t${ }}\fi}

\def\eps{\relax\ifmmode\epsilon\else{$\epsilon${ }}\fi}
\def\ee{\relax\ifmmode{e^+e^-}\else{${e^+e^-}$}\fi}
\def\qq{\relax\ifmmode{q\overline{q}}\else{$q\overline{q}${ }}\fi}

\newskip\humongous \humongous=0pt plus 1000pt minus 1000pt
\def\caja{\mathsurround=0pt}
\def\eqalign#1{\,\vcenter{\openup1\jot
\caja   \ialign{\strut \hfil$\displaystyle{##}$&$
\displaystyle{{}##}$\hfil\crcr#1\crcr}}\,}

\newif\ifdtup

\def\eqal2#1{\,\vcenter{\openup1\jot
\caja   \ialign{\strut \hfil$\displaystyle{##}$&\hfil$
\displaystyle{{}##}$\hfil &$
\displaystyle{{}##}$\hfil\crcr#1\crcr}}\,}
\def\app#1#2#3{{\em Acta Phys.\ Polon.}~{\underline {#1}} (19#3) #2}

\def\ib#1#2#3{{\em ibid.}~\underline{#1} (19#3) #2}
\def\ijmp#1#2#3{{\em Int.\ J.\ Mod.\ Phys.}~{\underline {#1}} (19#3) #2}

\def\np#1#2#3{{\em Nucl.\ Phys.}~\underline{B#1} (19#3) #2}
\def\npps#1#2#3{{\em Nucl.\ Phys.\ Proc.\ Suppl.}~\underline{#1} (19#3) #2}
\def\pl#1#2#3{{\em Phys.\ Lett.}~\underline{#1B} (19#3) #2}
\def\pr#1#2#3{{\em Phys.\ Rev.}~\underline{D#1} (19#3) #2}

\def\prl#1#2#3{{\em Phys.\ Rev.\ Lett.}~\underline{#1} (19#3) #2}

\def\spj#1#2#3{{\em Sov.\ Phys.\ JETP}\/~\underline{#1} (19#3) #2}
\def\zp#1#2#3{{\em Zeit.\ Phys.}~\underline{C#1} (19#3) #2}
 \def\cite#1{[\ref{#1}]}
 \def\citd#1#2{[\ref{#1},\ref{#2}]}
 \def\citt#1#2#3{[\ref{#1},\ref{#2},\ref{#3}]}
 
 \def\citm#1#2{[\ref{#1}--\ref{#2}]}
\def\cF{{\cal{F}}}
\def\a0{\bar\alpha_0}
\def\ae{\alpha_{\mbox{\scriptsize eff}}}
\def\aPT{\as^{\mbox{\scriptsize PT}}}

\def\mI{\mu_{\mbox{\tiny I}}}
\def\mR{\mu_{\mbox{\tiny R}}}
\def\re#1{(\ref{#1})}

\begin{document}
\begin{titlepage}
\begin{flushright}
Cavendish-HEP-96/5\\
hep-ph/9704297
\end{flushright}              
\vspace*{\fill}
\begin{center}
{\Large \bf Power Corrections to Event Shapes\\[1ex]
in Deep Inelastic Scattering\footnote{Research supported in
part by the U.K. Particle Physics and Astronomy Research Council and
by the EC Programme ``Training and Mobility of Researchers", Network
``Hadronic Physics with High Energy Electromagnetic Probes", contract
ERB FMRX-CT96-0008.}}
\end{center}
\par \vskip 5mm
\begin{center}
        M.\ Dasgupta and B.R.\ Webber \\
        Cavendish Laboratory, University of Cambridge,\\
        Madingley Road, Cambridge CB3 0HE, U.K.\\
\end{center}
\par \vskip 2mm
\begin{center} {\large \bf Abstract} \end{center}
\begin{quote}
We investigate the power-suppressed corrections to the mean
values of various quantities that characterise the shapes
of final states in deep inelastic lepton scattering. Our method
is based on an analysis of one-loop Feynman graphs containing a
massive gluon, which is equivalent to the evaluation of leading 
infrared renormalon contributions. As in $\ee$ annihilation,
we find that the leading corrections are proportional to $1/Q$.
We give quantitative estimates based on the hypothesis
of a universal low-energy effective coupling.
\end{quote}
\vspace*{\fill}
\begin{flushleft}
     Cavendish--HEP--96/5\\
     April 1997
\end{flushleft}
\end{titlepage}

\mysection{Introduction}
The study of final-state properties in deep inelastic lepton
scattering (DIS) has received a great impetus from the increasing
quantity and kinematic range of the HERA data. 
The determination of the strong coupling $\as$ from final-state
properties in DIS is an attractive possibility because of the
relative simplicity of the lepton-hadron interaction, combined with
the wide range of dynamical scales available in a single experiment
at a single beam energy. It is expected that the scale for $\as$
will be set primarily by the lepton-hadron momentum transfer-squared
$Q^2$, which can range from zero to 10$^5$ GeV$^2$ at HERA.  Thus
there should be a wide region in which the value and running of
$\as(Q^2)$ can be observed with good precision.

One possible method for $\as$ determination is the measurement of
jet fractions \cite{jetexp}, defined according to one of the several
available infrared-safe jet algorithms \citd{jetalg}{ktalg}. By definition,
jet rates defined by an infrared-safe algorithm can be computed
in perturbation theory, and next-to-leading-order calculations
are now available \citd{MZ}{CatSey}. The $\as$ values obtained by
comparing jet rates with HERA data are consistent with
those found in other processes, and in particular they show the
expected decrease with increasing $Q^2$.

In the present paper we consider a different set of DIS final-state
observables which can be used to determine $\as$. These are the
various {\em event shape variables} which can be defined in analogy
with those used in the study of $\ee$ annihilation final states.
In $\ee$ physics, event shapes have been found to be a useful
tool for testing QCD and measuring $\as$. They can be defined
so as to be sensitive to different aspects of QCD dynamics
(e.g.\ the longitudinal or transverse development of jets)
are subject to different non-perturbative `hadronization'
corrections. Thus $\as$ determinations from a variety of
event shapes complement those from jet rates and give an
indication of the systematic uncertainties due to
non-perturbative effects. The same considerations make it
important to calculate and measure event shapes in DIS.

Another reason to measure event shapes is that there are
new theoretical ideas about non-perturbative corrections
to them \citm{hadro}{KorSte}, which provide constraints on $\as$
determinations from event shapes and are interesting to test in
their own right. By looking at the behaviour of the QCD
perturbation series in high orders, one can identify
unsummable, factorially divergent sets of contributions
(infrared renormalons \cite{renormalons}) which indicate
that non-perturbative power-suppressed corrections must
be included. The $Q^2$-dependence of the leading correction
to a given quantity can be inferred, and by making further
universality assumptions one may also estimate its magnitude.
Tests of these ideas provide information on the transition
from the perturbative to the non-perturbative regime in QCD.
In particular, one can investigate the possibility that an
approximately universal low-energy effective coupling may be
a useful phenomenological concept \citt{DokWeb}{BPY}{alfaeff}.

Such an approach has been applied with some success to
$\ee$ event shapes \citt{DokWeb}{BPY}{shapexp} and fragmentation
functions \cite{fragren}, and to DIS structure
functions \citd{BPY}{DISren}. In the present paper we extend it to
event shapes in DIS \cite{WParis}.  We find that, as in $\ee$
annihilation, the leading
power corrections to these quantities are typically proportional
to $1/Q$.  The hypothesis that they are related to a universal
low-energy effective coupling implies that their magnitudes
are given by a single non-perturbative parameter. We give
quantitative estimates based on the value of this parameter
derived from $\ee$ data.

In the following Section we explain how the DIS event shape
variables that we compute are defined.  In Sect.~3 we give
the leading-order perturbative predictions for these quantities.
To determine $\as$, one needs the predictions in next-to-leading
order, which are not yet available.  However, the leading-order
calculation provides a useful guide to the relative importance
of the power-suppressed corrections, which we estimate in
Sect.~4 using the method of Ref.~\cite{BPY}. We explain how
these estimates can be refined and combined with the
next-to-leading predictions when they become available.
Finally, our results are summarized briefly in Sect.~5.

\mysection{Event shape variables in DIS}\label{sec_shapes}
A complication in DIS, absent from $\ee$ annihilation,
is the presence in the final state of the remnant of the
initial-state hadron, i.e.\ the constituents that did not
participate in the hard scattering of the lepton.  It is
expected that the fragmentation of the remnant will be
dominated by soft, non-perturbative physics. While of
interest for studying the hadronization process,
the remnant fragmentation is not so useful for $\as$
determinations, and therefore we concentrate here on
aspects of event shapes that are not sensitive to it.
This is conveniently done by looking at the final state
in the {\em Breit frame of reference} \citt{SWZ}{PeRu}{GDKT}.

We consider the deep inelastic scattering of a lepton of
momentum $l$ from a nucleon of momentum $P$, with momentum
transfer $q$.  The main kinematic variables are $Q^2 = -q^2$,  
the Bjorken variable $x=Q^2/2P\cdot q$ and
$y=P\cdot q/P\cdot l\simeq Q^2/x s$, $s$ being the total
c.m.\ energy squared. Then the Breit frame is the rest-frame
of $2xP+q$. In this frame the momentum transfer $q$ is purely
spacelike, and we choose to align it along the $+z$ axis:
\beq\label{Pqmoms}
P = \half Q (1/x,0,0,-1/x) \;,\;\;\;\;
q = \half Q (0,\,0,0,2)\;.
\eeq

To a good approximation, the fragmentation products of the remnant
will be moving in directions close to that of the incoming nucleon,
i.e.\ they will remain in the `remnant hemisphere' $H_r$ ($p_z<0$).
On the other hand the products of the hard lepton scattering
will tend to be found in the `current hemisphere' $H_c$
($p_z>0$). In fact in the parton model the scattered parton
moves along the current ($+z$) axis with momentum
$xP+q = \half Q (1,0,0,1)$.
Thus in the parton model the current hemisphere looks like
one hemisphere of the final state in $\ee$ annihilation at
centre-of-mass energy $Q$. Fragmentation studies have
shown that this similarity is indeed manifest in hadron
spectra and multiplicities \cite{frag}.  This makes it natural
to define event shape variables in close analogy to those
for $\ee$ annihilation, but limited to particles $a$ appearing
in the current hemisphere, $a\in H_c$. 

We can now construct infrared-safe quantities that characterize the
shape of the event defined in this way.  Perhaps the
simplest is the {\em current jet thrust} \cite{SWZ}
\beq\label{TQdef}
T_Q = 2\sum_{a\in H_c} \bm{p}_a\cdot\bm{n}\Big/Q
\eeq
where $\bm{n}$ represents the unit 3-vector along the current
direction (the $+z$ axis, in our convention).
The subscript $Q$ indicates that $T$ is normalized
to $\half Q$. Alternatively we may normalize
to the total energy in the current hemisphere,
\beq\label{TEdef}
T_E = \sum_{a\in H_c} \bm{p}_a\cdot\bm{n}\Big/\sum_{a\in H_c} E_a\;.
\eeq
Both of these quantities are equal to unity in the Born approximation,
and their deviation from this value measures the longitudinal
development of the current jet. It will therefore be convenient
to study instead the quantities $\tau_Q=1-T_Q$ and $\tau_E=1-T_E$,
which vanish in the Born approximation.

It is kinematically possible for the Breit
frame current hemisphere to be empty. In that case, taken literally,
Eq.~\re{TQdef} implies that $T_Q=0$, hence $\tau_Q=1$, while
Eq.~\re{TEdef} leaves $\tau_E$ undefined.  For consistency with
the other event shapes defined below, we instead define
$\tau_Q=\tau_E=0$ when the current hemisphere is empty. 

Similarly we can define the {\em current jet broadening} \cite{CTW}
\beq\label{BQdef}
B_Q = \sum_{a\in H_c} |\bm{p}_a\times\bm{n}|\Big/Q\;,
\eeq
or
\beq\label{BEdef}
B_E = \half\sum_{a\in H_c} |\bm{p}_a\times\bm{n}|\Big/\sum_{a\in H_c} E_a\;,
\eeq
which emphasizes the transverse development of the jet.

Both the thrust and the broadening are defined here with
respect to the current direction $\bm{n}$. Two quantities which
measure the jet development independent of direction (apart from
the restriction to particles in the current hemisphere) are the
scaled {\em current jet mass}
\beq\label{RQdef}
\rho_Q = \left(\sum_{a\in H_c} p_a\right)^2\Big/Q^2
\eeq
and the {\em $C$-parameter} \cite{ERT}
\beq\label{CQdef}
C_Q = 3(\lambda_1\lambda_2+\lambda_2\lambda_3+\lambda_3\lambda_1)
\eeq
where $\lambda_{1,2,3}$ are the eigenvalues of the linearized momentum
tensor
\beq
\Theta^{ij} = 2 \sum_{a\in H_c} \left(\bm{p}_a^i\bm{p}_a^j/|\bm{p}_a|\right)
\Big/Q\;.
\eeq
Again, we may alternatively define quantities $\rho_E$ and
$C_E$, in which $Q$ is replaced by twice the total energy
in the current hemisphere, so that
\beq\label{RCEdef}
\rho_E/\rho_Q = C_E/C_Q = Q^2\Big/\left(2\sum_{a\in H_c} E_a\right)^2\;.
\eeq

\begin{figure}\begin{center}
\epsfig{figure=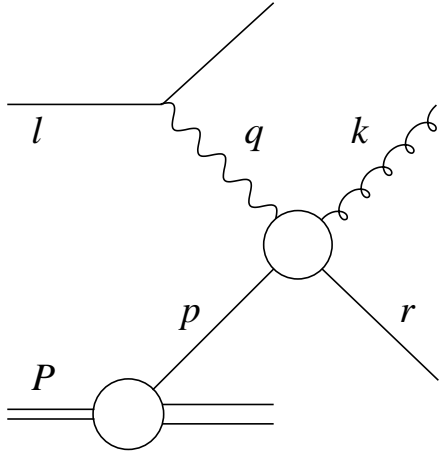, height=6.0cm}
\caption{Jet production in deep inelastic scattering.}
\label{fig_disjet}
\end{center}\end{figure}

\mysection{Leading-order perturbation theory}
At first order in $\as$, up to two final-state partons can be
emitted in the hard lepton-parton subprocess, as illustrated in
Fig.~\ref{fig_disjet}. The momentum of the struck parton is
$p = xP/\xi$ ($x<\xi<1$) and we define $z=P\cdot r/P\cdot q$ ($0<z<1$).
 
The differential cross section is 
\beq\label{diffcs}
\frac{d^3\sigma}{dx dQ^2 dz} = \frac{2\pi\alpha^2}{Q^4}
\left\{\left[1+(1-y)^2\right] F_T(x,z)+2(1-y) F_L(x,z)\right\}\;.
\eeq
The generalized transverse and longitudinal structure functions
$F_T(x,z)=2F_1(x,z)$ and $F_L(x,z) = F_2(x,z)/x - 2F_1(x,z)$ are
of the form (for $z<1$)
\beq
F_i(x,z) = \frac{\as}{2\pi} \int_x^1 \frac{d\xi}{\xi}
\left[C_F C_{i,q}(\xi,z) q(x/\xi) + T_f C_{i,g}(\xi,z)g(x/\xi)\right]
\eeq
where
\beq
q(x) = \sum_{j=1}^f e_j^2 \left[q_j(x)+\bar q_j(x)\right]\;,
\;\;\;\;\;\; T_f = T_R\sum_{j=1}^f e_j^2
\eeq
for $f$ active quark flavours, $C_F=4/3$, $T_R=1/2$ and \cite{PeRu}
\beq\label{C0}\eqalign{
C_{T,q}(\xi,z) &= \frac{\xi^2+z^2}{(1-\xi)(1-z)}+2\xi z+2 \cr
C_{L,q}(\xi,z) &= 4\xi z \cr
C_{T,g}(\xi,z) &= \left[\xi^2+(1-\xi)^2\right]\frac{z^2+(1-z)^2}{z(1-z)}\cr
C_{L,g}(\xi,z) &= 8\xi(1-\xi)\;.}
\eeq

In the Breit frame $P$ and $q$ are given by Eq.~\re{Pqmoms}
and we can write
\beq\label{moms}\eqalign{
p &= \half Q (1/\xi,0,0,-1/\xi) \cr
r &= \half Q (z_0,z_\perp,0,z_3) \cr
k &= \half Q (\bar z_0,-z_\perp,0,\bar z_3)}
\eeq
where
\beq\label{zs0}\eqalign{
z_0 &= 2z-1+(1-z)/\xi\cr
z_3 &= 1-(1-z)/\xi\cr
\bar z_0 &= 1-2z+z/\xi\cr
\bar z_3 &= 1-z/\xi\cr
z_\perp &= 2\sqrt{z(1-z)(1-\xi)/\xi}\;.}
\eeq

\begin{figure}\begin{center}
\epsfig{figure=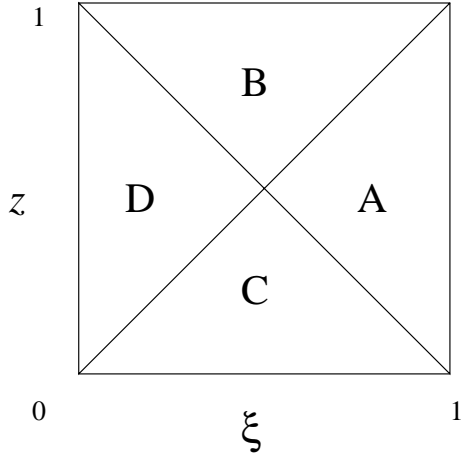, height=6.0cm}
\caption{Phase space region for jet production in deep inelastic scattering.}
\label{fig_disregion}
\end{center}\end{figure}

We can distinguish four subregions of phase space,
as illustrated in Fig.~\ref{fig_disregion}:

A: both produced parton momenta $k,r$ in the current
hemisphere ($z_3,\bar z_3>0$);

B: only parton momentum $r$ in the current
hemisphere ($z_3 >0,\;\bar z_3<0$);

C: only parton momentum $k$ in the current
hemisphere ($z_3 <0,\;\bar z_3>0$);

D: no produced parton momenta in the current
hemisphere ($z_3, \bar z_3 <0$).

In leading order the event shape variables defined in
Sect.~\ref{sec_shapes} are given in these regions
by Table~1. By definition they are all zero in region D.
By construction, they all vanish in the soft and/or
collinear limits $\xi, z\to 1$. Note that $\rho$ and
$C$ also vanish throughout regions B and C.
 
\begin{table}[t]
\caption{Event shape variables $S(\xi,z)$ in leading order.}
\begin{center}
\begin{tabular}{|c|c|c|c|}\hline
$S$ & A & B & C \\ \hline
 & & & \\
$\tau_Q$ & $(1-\xi)/\xi$ & $1-z_3$ & $1-\bar z_3$ \\
 & & & \\
$\tau_E$ & $2(1-\xi)$ & $1-z_3/z_0$ & $1-\bar z_3/\bar z_0$ \\
 & & & \\
$B_Q$ & $z_\perp$ & $z_\perp/2$ & $z_\perp/2$ \\
 & & & \\
$B_E$ & $\xi z_\perp$ & $z_\perp/2z_0$ & $z_\perp/2\bar z_0$ \\
 & & & \\
$\rho_Q$ & $(1-\xi)/\xi$ & 0 & 0 \\
 & & & \\
$\rho_E$ & $\xi(1-\xi)$ & 0 & 0 \\
 & & & \\
$C_Q$ & $3(2\xi-1)^2 z_\perp^2/\xi^2 z_0\bar z_0$ & 0 & 0 \\
 & & & \\
$C_E$ & $3(2\xi-1)^2 z_\perp^2/ z_0\bar z_0$ & 0 & 0 \\
 & & & \\
\hline \end{tabular}
\end{center}\end{table}

At any particular values of $x$ and $Q^2$, the mean value of
a shape variable $S$ is now given in leading order by
\beq\label{VEVSdef}
\VEV{S} = \frac{2\pi\alpha^2}{Q^4} \int_0^1 dz 
\left\{\left[1+(1-y)^2\right] F^{(S)}_T(x)+2(1-y) F^{(S)}_L(x)\right\}
\bigg/ \frac{d^2\sigma_0}{dx dQ^2} 
\eeq
where
\beq
F^{(S)}_i(x) = \frac{\as}{2\pi}\int_x^1\frac{d\xi}{\xi}\int_0^1 dz S(\xi,z)
\left[C_F C_{i,q}(\xi,z) q(x/\xi) + T_f C_{i,g}(\xi,z)g(x/\xi)\right]
\eeq
and the denominator is the differential cross section evaluated
in Born approximation,
\beq\label{Borncs}
\frac{d^2\sigma_0}{dx dQ^2} = \frac{2\pi\alpha^2}{Q^4}
\left[1+(1-y)^2\right] q(x)\;.
\eeq

We discuss the numerical values of the leading-order predictions
together with the power corrections in the following Section.

\mysection{Power corrections}
Our estimate of the leading power corrections to the perturbative
results given above is based on the approach of Ref.~\cite{BPY}.
Non-perturbative effects at long distances are assumed to
give rise to a modification $\delta\ae(\mu^2)$ in the QCD
effective coupling at low values of the scale $\mu^2$. The
effect on some observable $F$ is then given by a
{\it characteristic function} $\cF(x,\eps)$, as follows:
\beq\label{deltaF}
\delta F(x,Q^2) = \int_0^\infty \frac{d\mu^2}{\mu^2}\, 
\delta\ae(\mu^2)\dot\cF(x,\eps=\mu^2/Q^2)
\eeq
where
\beq\label{dotFdef}
\dot\cF(x,\eps) \equiv -\eps\frac{\partial}{\partial\eps}\cF (x,\eps)\;.
\eeq
The characteristic function is obtained by computing the
relevant one-loop graphs with a non-zero gluon mass
$\mu=Q\sqrt\eps$ \citd{hadro}{BBB}.

Arbitrary finite modifications of the
effective coupling at low scales would generally
introduce power corrections of the form  $1/\mu^{2p}$ into the
ultraviolet behaviour of the running coupling $\as$ itself.
Such a modification would destroy the basis of the operator
product expansion \cite{SVZ}.  One must therefore require
that at least the first few integer moments of the coupling
modification should vanish:
\beq\label{vanish}
 \int_0^\infty \frac{d\mu^2}{\mu^2}
\>\left(\mu^2\right)^p \delta\ae(\mu^2)\>=\>0\>; \quad p=1,\ldots,p_{\max}\,. 
\eeq
The upper bound $p_{\max}$ could be set by instanton--anti-instanton
contributions ($p_{\max}\sim 9$). The constraint \re{vanish} means
that only those terms in the small-$\eps$ behaviour of the characteristic
function that are {\em non-analytic} at $\eps =0$ will lead to
power-behaved non-perturbative contributions. These are just the
the terms that give rise to infrared renormalons in perturbation
theory \cite{BBB}.

\begin{figure}\begin{center}
\epsfig{figure=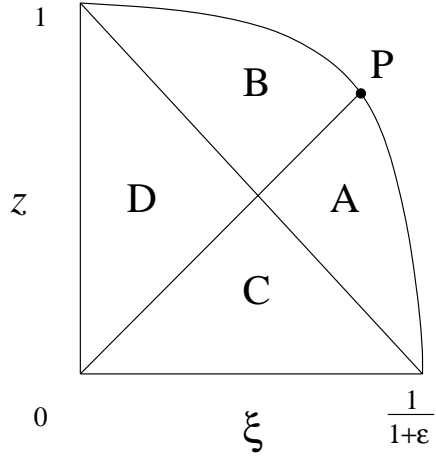, height=6.0cm}
\caption{Phase space region with gluon mass-squared $\mu^2=\eps Q^2$.}
\label{fig_disregeps}
\end{center}\end{figure}

For gluon mass-squared $\mu^2 =\eps Q^2$ the quark coefficient
functions in Eq.~\re{C0} become
\beq\label{Ceps}\eqalign{
C_{T,q}(\xi,z,\eps) &=
\frac{(1-z)(1-\xi)+2\xi z(1-z)^2-\xi\eps}{(1-z-\xi\eps)^2}
+\frac{2\xi z(1-\eps)}{(1-z-\xi\eps)(1-\xi)}
+\frac{(1-z)(1-\xi)-\xi\eps}{(1-\xi)^2}\cr
C_{L,q}(\xi,z,\eps) &=
\frac{4\xi z(1-z)^2}{(1-z-\xi\eps)^2}\;.}
\eeq
Processes involving an incoming gluon are not expected to
give terms that are non-analytic at $\eps=0$, and therefore
we do not consider them as a source of power corrections.
The kinematic variables that give the momenta according to
Eq.~\re{moms} are now
\beq\label{zseps}\eqalign{
z_0 &= 2z-1+(1-z)/\xi-\eps\cr
z_3 &= 1-(1-z)/\xi+\eps\cr
\bar z_0 &= 1-2z+z/\xi+\eps\cr
\bar z_3 &= 1-z/\xi-\eps\cr
z_\perp &= 2\sqrt{z(1-z)(1-\xi)/\xi-\eps z}
\;.}\eeq
Thus the phase space region is now $0 < z < 1-\eps \xi/(1-\xi)$, as
illustrated in Fig.~\ref{fig_disregeps}. The regions A,\ldots D
defined above in terms of the signs of $z_3$ and $\bar z_3$ are
as indicated.

The corresponding characteristic function for the
mean value of some event shape variable $S$ is given by
Eq.~\re{VEVSdef} with $F^{(S)}_i(x)$ replaced by
$(C_F/2\pi) \cF_i^{(S)}(x,\eps)$ where
\beq\label{cFtau}
\cF_i^{(S)}(x,\eps) =  \int_{x}^1\frac{d\xi}{\xi}
\int_0^1 dz\,S(\xi,z,\eps)C_{i,q}(\xi,z,\eps)
\Theta(1-z-\xi+\xi z-\eps \xi)\,q(x/\xi)\;.
\eeq
Note that for brevity we have extracted the overall factor of $C_F/2\pi$.
The expressions for the shape variables $S(\xi,z,\eps)$ are as given
in Table~1, but with the kinematic variables now given by Eqs.~\re{zseps}
instead of Eqs.~\re{zs0}.

Finding the leading non-analytic term in the behaviour of the integral
\re{cFtau} as $\eps\to 0$, differentiating as instructed in
Eq.~\re{dotFdef}, and inserting the result in Eq.~\re{deltaF},
we obtain the corresponding predicted power correction.  
Generally speaking, the small-$\eps$ behaviour of the characteristic
function is dominated by the region around the boundary point P in
Fig.~\ref{fig_disregeps}, and therefore the leading power correction
is independent of whether we normalize the shape variable to $Q/2$ or
to the energy in the current hemisphere. 

Following Ref.~\cite{BPY}, we may express the magnitudes of power
corrections in terms of the moment integrals
\beq\label{adefs}
A_{2p} = \frac{C_F}{2\pi}
\int_0^\infty \frac{d\mu^2}{\mu^2}\,\mu^{2p}\,\delta\ae(\mu^2)\;,
\eeq
which vanish for integer $p$, and their $p$-derivatives
\beq\label{apdefs}
A'_{2p} = \frac{C_F}{2\pi}
\int_0^\infty \frac{d\mu^2}{\mu^2}\,\mu^{2p}\,\log\mu^2\,\delta\ae(\mu^2)\;,
\eeq
which are in general non-vanishing for any $p$. The leading corrections
to the event shapes we are considering correspond to $p=\half$, and
therefore, on the assumption that $\delta\ae$ is universal, they can
all be expressed in terms of the two non-perturbative parameters
$A_1$ and $A'_1$.  Studies of event shapes in $\ee$ annihilation
suggest that $A_1\simeq 0.25$ GeV \cite{BPY}, with $A'_1$ as yet
undetermined.

As an alternative representation of the magnitudes of
power corrections, we may adopt the approach of Ref.~\cite{DokWeb}
and express them directly in terms of moments of $\ae$ over the
infrared region. We substitute for $\delta\ae$ in Eq.~\re{adefs}
\beq
\delta\ae(\mu^2) \simeq \ae(\mu^2) - \aPT(\mu^2)\;,
\eeq
where $\aPT$ represents the expression for $\as$
corresponding to the part already included in the
perturbative prediction.  As discussed in Ref.~\cite{DokWeb},
if the perturbative calculation is carried out to second
order in the \MSbar\ renormalization scheme, with renormalization
scale $\mR^2$, then we have
\beq
\aPT(\mu^2) = \as(\mR^2) + [b\ln(\mR^2/\mu^2)+k]\,\as^2(\mR^2) 
\eeq
where ($C_A=3$)
\beq
b = \frac{11 C_A-2f}{12\pi}\;,\;\;\;\;
k= \frac{(67-3\pi^2)C_A-10f}{36\pi}\;.
\eeq
The constant $k$ comes from a change of scheme from \MSbar\ to
the more physical scheme \cite{CMW} in which $\ae$ is defined.
Then above some infrared matching scale $\mI$ we assume that
$\ae(\mu^2)$ and $\aPT(\mu^2)$ approximately coincide, so that
\beq\label{A1exp}\eqalign{
A_1 &\simeq \frac{C_F}{2\pi}
\int_0^{\mI^2} \frac{d\mu^2}{\mu^2}\,\mu\,\left(\ae(\mu^2)-
\as(\mR^2) - [b\ln(\mR^2/\mu^2)+k]\,\as^2(\mR^2)\right)\cr
&= \frac{C_F}{\pi}\,\mI\,\left(\a0(\mI)-
\as(\mR^2) - [b\ln(\mR^2/\mI^2)+k+2b]\,\as^2(\mR^2)\right)\;,}
\eeq
where
\beq
\a0(\mI) \equiv \frac{1}{\mI}\int_0^{\mI} \ae(\mu^2)\,d\mu\;.
\eeq
Thus in this notation the value of $A_1$ determines the
average value of the effective coupling below the matching
scale $\mI$. The dependence of $\a0$ on $\mI$ is partially
compensated by the $\mI$-dependence of the other terms on
the right-hand side of Eq.~\re{A1exp}. The dependence on the
renormalization scale $\mR^2$ should help to compensate the
scale dependence of the perturbative part.  Notice that if
we take $\mR^2\propto Q^2$ then $A_1$ has a logarithmic dependence
on $Q^2$.  In general we do expect `power' corrections to have
additional logarithmic $Q^2$-dependence (anomalous dimensions),
but this cannot yet be calculated reliably for event shapes.
 
In Ref.~\cite{DokWeb} it was found that the formula \re{A1exp}
with $\mR^2=Q^2$, $\mI=2$ GeV and $\a0$(2 GeV) = 0.52
gave good agreement with $\ee$ event shape data. Similar
results were obtained in Ref.~\cite{shapexp}. 

\subsection{Current jet thrust}
In the case of the shape variables $\tau_Q$ or $\tau_E$, the behaviour of
the expression \re{cFtau} for the transverse contribution
$\cF_T^{(\tau)}(x,\eps)$ as $\eps\to 0$ is found to be of the form
\beq\label{cFtep}
\cF_T^{(\tau)}(x,\eps) \sim \cF_T^{(\tau)}(x,0)
-8\sqrt{\eps}\,q(x)\;,
\eeq
while the longitudinal part $\cF_L^{(\tau)}$ is less singular at $\eps=0$.
Thus from Eqs.~\re{VEVSdef}, \re{deltaF} and \re{adefs}
we obtain the leading non-perturbative contribution
\beq\label{dVEVt}
\delta\VEV{\tau} \sim 4\frac{A_1}{Q}\;.
\eeq
The behaviour \re{cFtep} at small $\eps$ follows from the fact that
the derivative $\dot\cF_T^{(\tau)}$ is dominated by the phase space
boundary $z=1-\eps \xi/(1-\xi)$:
\beq\label{cFt0}
\dot\cF_T^{(\tau)}(x,\eps) \sim \eps \int_{x}^1 d\xi
\int_0^1 dz\,\tau(\xi)C_{i,q}(\xi,z,0)
\delta(1-z-\xi+\xi z-\eps \xi)\,q(x/\xi)\;.
\eeq
From Table~1 and Eq.~\re{zseps}, on this boundary we have
\beq\eqalign{
\tau(\xi,z,0) & = (1-\xi)/\xi \;\;\mbox{for}\;\xi>\xi_P\;,\cr
 & = (1-z)/\xi \;\;\mbox{for}\;\xi<\xi_P}
\eeq
where $\xi_P=1/(1+\sqrt{\eps})$. Thus
\beq\label{cFtapp}
\dot\cF_T^{(\tau)}(x,\eps) \sim \int_{x}^{\xi_P}\frac{d\xi}{\xi}\,
\eps\,\frac{1+\xi^2}{(1-\xi)^2}\,q\left(\frac{x}{\xi}\right) 
+ \int_{\xi_P}^1\frac{d\xi}{\xi}\,
\frac{1+\xi^2}{\xi}\,q\left(\frac{x}{\xi}\right) 
\>\sim\> 4\sqrt{\eps}\,q(x)\;,
\eeq
in agreement with Eq.~\re{cFtep}.

\begin{figure}\begin{center}
\epsfig{figure=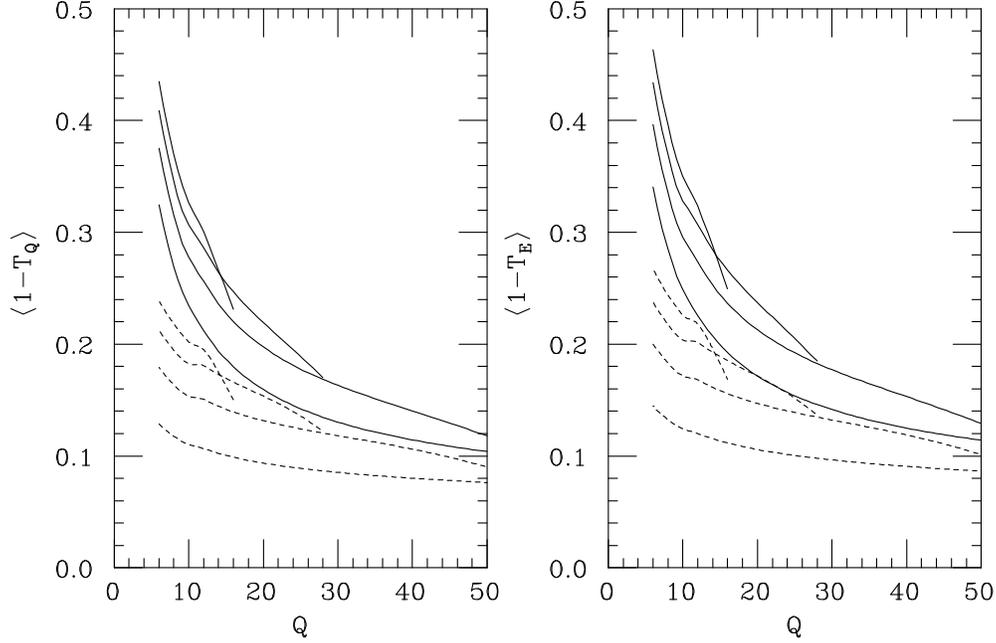, height=8.0cm}
\caption{Predictions for the mean value of the current jet thrust
in deep inelastic scattering.
Left- and right-hand plots are for the two definitions
(2.2) and (2.3), respectively.
Dashed: leading-order perturbation theory.
Solid: leading order plus leading power correction.
In each case the four curves (top to bottom) are for
$x=0.003, 0.01,0.03,0.10$.}
\label{fig_shapdis_T}
\end{center}\end{figure}

Numerical predictions for the mean value of the current jet thrust
in $ep$ scattering at $\sqrt s =296$ GeV
are shown in Fig.~\ref{fig_shapdis_T} as a function of $Q$
for various values of $x$. The MRS A$'$ parton distributions \cite{MRSA}
were used, with the corresponding value
$\Lambda^{(4)}_{\overline{\mbox{\scriptsize MS}}} = 231$ MeV in
the two-loop expression for $\as(Q^2)$. The leading-order perturbative
predictions given by Eq.~\re{VEVSdef} are shown by the dashed curves.
For the power correction coefficient $A_1$ we used Eq.~\re{A1exp}
with $\mR^2=Q^2$, $\mI=2$ GeV and $\a0$(2 GeV) = 0.52, as in
the fits to $\ee$ data, but we omitted the term of order $\as^2$
because we are combining with only a first-order perturbative
calculation in this paper. When the higher-order prediction
becomes available, the ${\cal O}(\as^2)$ term in Eq.~\re{A1exp}
should be included when estimating the power correction to it.

The resulting overall predictions are shown by the solid curves.
We see that the estimated power correction is substantial,
20-30\% at $Q=50$ GeV and dominating below 15 GeV. There is
significant $x$-dependence in the perturbative prediction,
and large-$x$ data ($x>0.01$) are required to cover the
region where the power correction is under control. With
sufficient data in the range $Q=15-50$ GeV, however, it
should be possible to perform a two-parameter fit to
determine $\as$ and $\a0$ from the average current jet thrust.

\subsection{Current jet broadening}
For the jet broadening $B_Q$ or $B_E$ we find a slightly different
behaviour at small $\eps$, namely
\beq\label{cFbep}
\cF_T^{(B)}(x,\eps) \sim \cF_T^{(B)}(x,0)
+8\sqrt{\eps}\,(\ln\eps + c)\,q(x)\;,
\eeq
where $c$ is a constant (probably $x$-independent) which we cannot
determine reliably.  This implies a non-perturbative correction of
the form
\beq
\delta\VEV{B} \sim 4\frac{A_1}{Q}(\ln Q^2-c-2) - 4\frac{A'_1}{Q}\;.
\eeq
Since we do not know the value of $c$, we may as well absorb all
the non-logarithmic terms into an unknown scale, $Q_0$:
\beq\label{dVEVB}
\delta\VEV{B} = 8\frac{A_1}{Q}\ln(Q/Q_0)\;.
\eeq

We find that, in contrast to the situation for the thrust, the
result \re{cFbep} is only obtained when one includes the gluon
mass explicitly in the definition of the jet broadening. This is
because, unlike the thrust, the broadening vanishes on the
phase-space boundary $z_\perp=0$. If we neglect the gluon mass
in the definition, this is not the case and an expression
analogous to Eq.~\re{cFt0} is obtained, which gives
\beq
\dot\cF_T^{(B)}(x,\eps) \sim \sqrt{\eps} \int_{x}^{\xi_P}
\frac{d\xi}{1-\xi}\frac{\xi^2+1}{\xi}\,q\left(\frac{x}{\xi}\right)
+ 2 \sqrt{\epsilon}\int _{\xi_P}^{1/(1+\epsilon)}
\frac{d\xi}{1-\xi}\frac{\xi^2 + 1}{\xi}\,q\left(\frac{x}{\xi}\right)\;. 
\eeq
We then find
\beq
\dot\cF_T^{(B)}(\eps) \sim -3 \sqrt{\eps}\,(\ln\eps + c)\;,
\eeq
corresponding to a coefficient of 6 instead of 8 in Eq.~\re{dVEVB}.
Thus the correct form is obtained, but the full mass-dependence
must be retained to compute the coefficient.

\begin{figure}\begin{center}
\epsfig{figure=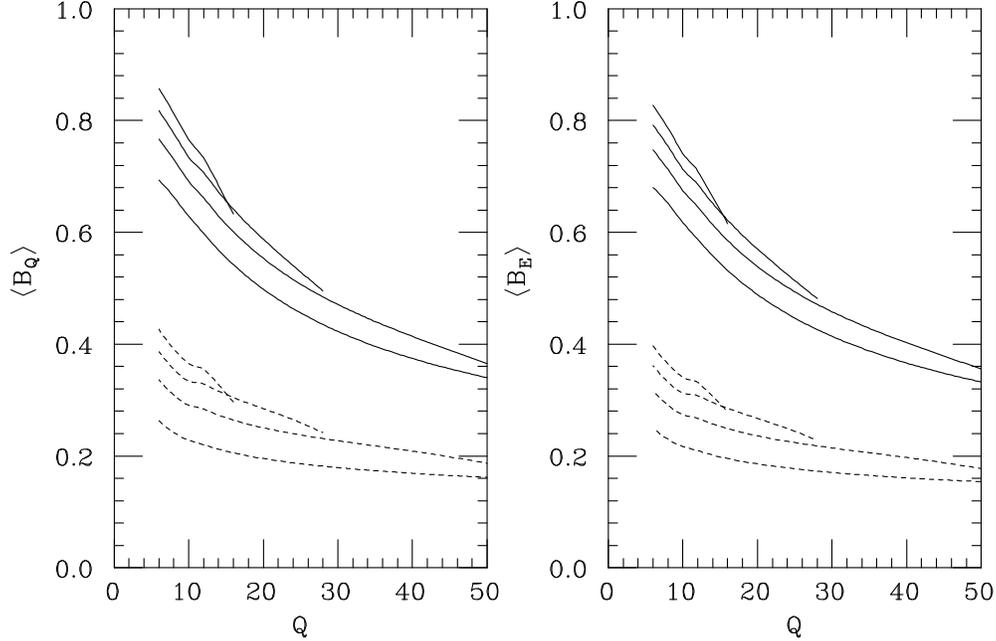, height=8.0cm}
\caption{Predictions for the mean value of the current jet broadening
in deep inelastic scattering.  Left- and right-hand
plots are for the two definitions (2.4) and (2.5), respectively.
Curves as in Fig.~3.}
\label{fig_shapdis_B}
\end{center}\end{figure}

The fact that the magnitude of the leading power correction to the jet
broadening is sensitive to the gluon mass-dependence in its definition
suggests to us that the prediction for this shape variable less reliable
than that for the thrust. As pointed out in Ref.~\cite{NaSey}, shape
variables are not fully inclusive with respect to the fragmentation
of the gluon: their values for the `decay products' of a timelike
virtual gluon are not necessarily equal to those for a `real'
gluon of equivalent mass.  In the case of the thrust, model studies
suggest that the numerical effect of this on the leading power
correction is small, but we expect it to be larger for variables
that depend explicitly on the gluon mass.

Numerical predictions for the current jet broadening at HERA are
shown in Fig.~\ref{fig_shapdis_B}, using the same parameter values
as before to compute $A_1$ and, for definiteness, $Q_0=\mI=2$ GeV
in Eq.~\re{dVEVB}.  We see that the resulting power corrections
are large. As we have stressed above, they are also more
uncertain in this case, suggesting that jet broadening in DIS
is not a good shape variable for $\as$ determinations.

\subsection{Current jet mass}
Next we consider the power corrections to the jet mass $\rho_Q$ or
$\rho_E$. We notice from Table 1 that there is no explicit gluon mass
dependence in the definition of these variables, and there is no
contribution outside the phase space region A.\footnote{There is
a contribution $\eps$ in region C, but since this is analytic it
does not contribute to the power correction.}
Inside this region we have in fact  $\rho_Q=\tau_Q$. Thus the
jet mass receives a contribution from the second integral only
in Eq.~\re{cFtapp}. This gives exactly one half of the correction
to the thrust and so one finds that
\beq 
\cF_T^{(\rho)}(x,\eps) \sim \cF_T^{(\rho)}(x,0)
-4\sqrt{\epsilon} \, q(x)
\eeq
at small $\eps$, which implies a non-perturbative correction
\beq
\delta \VEV{\rho} \sim 2 \frac{A_1}{Q}\;.
\eeq

\begin{figure}\begin{center}
\epsfig{figure=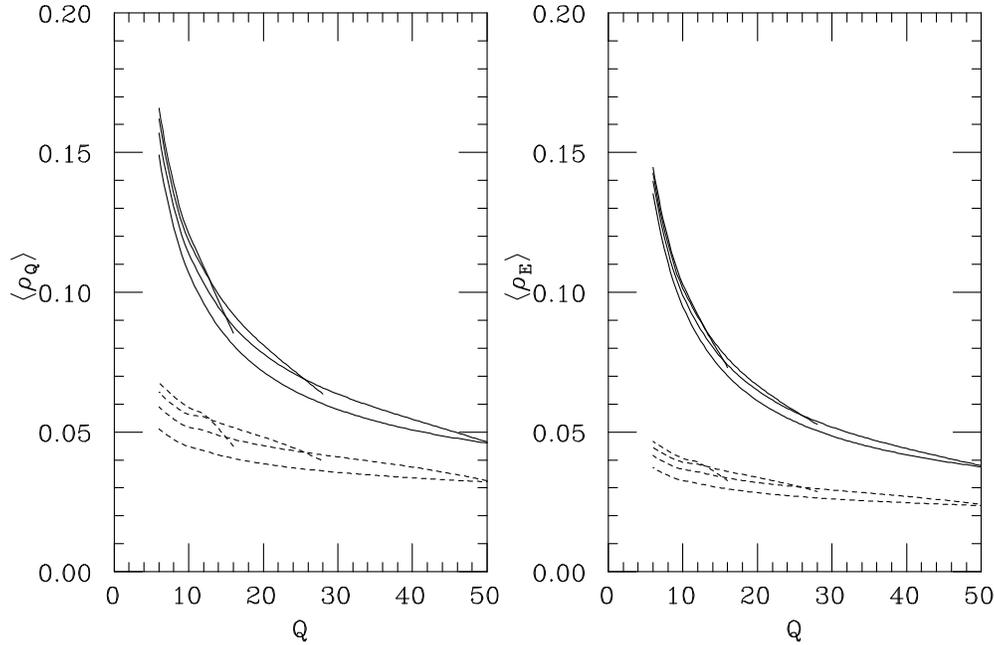, height=8.0cm}
\caption{Predictions for the mean value of the current jet mass
in deep inelastic scattering.  Left- and right-hand
plots are for the two definitions (2.6) and (2.9), respectively.
Curves as in Fig.~3.}
\label{fig_shapdis_R}
\end{center}\end{figure}

The numerical predictions for the current jet mass, shown in
Fig.~\ref{fig_shapdis_R}, suggest that this is a good variable
for $\as$ determinations. The power correction is somewhat
larger than that for the thrust, relative to the perturbative
prediction (cf.\ Fig.~\ref{fig_shapdis_T}), but there is less
$x$ dependence in the latter.

\subsection{\boldmath $C$-parameter}
Finally we compute the power correction to $C_Q$ or $C_E$.
Here again we see from Table~1 that that there is no
contribution outside the phase space region A.  However in this
case the variable, unlike the jet mass, does depend explicitly on the
gluon mass.  From a full evaluation retaining this mass dependence we
find the small-$\eps$ behaviour
\beq\label{cFCep}
\cF_T^{(C)}(x,\eps) \sim \cF_T^{(C)}(x,0)
-24\pi\sqrt{\eps}\,q(x)\;,
\eeq
corresponding to a leading non-perturbative contribution
\beq\label{dVEVC}
\delta\VEV{C} \sim 12\pi\frac{A_1}{Q}\;.
\eeq
If one uses the massless gluon expression for $C$, one obtains
\beq
\dot\cF_T^{(\tau)}(x,\eps)\sim 12 \eps \int_{1/(1+\sqrt{\eps})}^{1/(1+\eps)}
\frac{\xi^2 + 1}{\xi ^2}  \; \frac{(2\xi-1)^2 (1-\xi)}{1- \xi
+\eps (1-2\xi)}\; \frac{d \xi}{(1-\xi)^2 +\eps\xi (2\xi-1)}\sim 6\pi
\sqrt{\eps}\;,
\eeq
which corresponds to one-half of the full result. We would argue again
that this sensitivity to the gluon mass-dependence of the definition
suggests that the prediction for the $C$-parameter is less reliable
than that for the thrust and jet mass.

\begin{figure}\begin{center}
\epsfig{figure=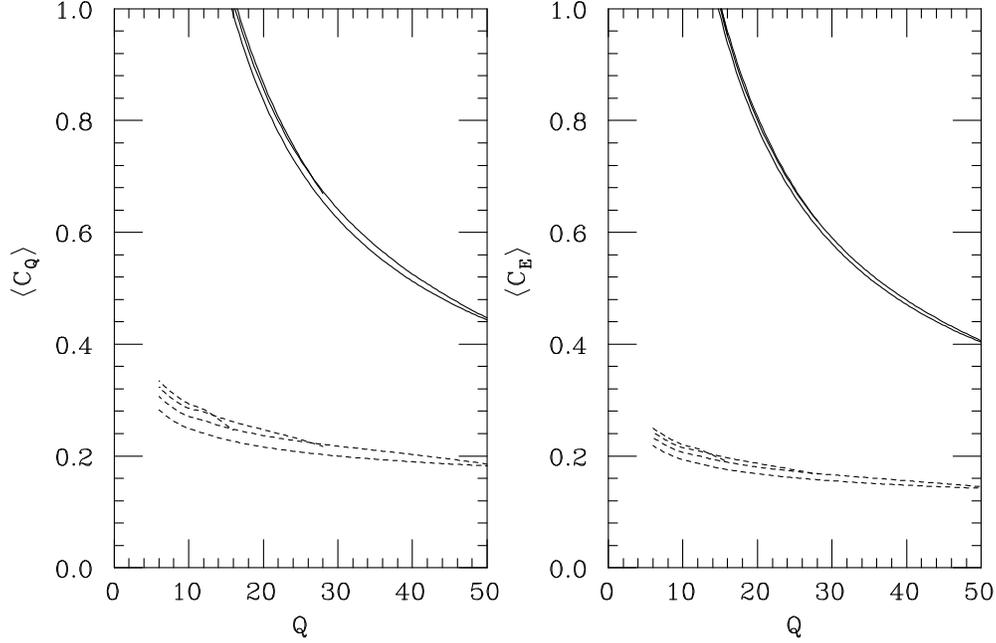, height=8.0cm}
\caption{Predictions for the mean value of the $C$-parameter
in deep inelastic scattering.  Left- and right-hand
plots are for the two definitions (2.7) and (2.9), respectively.
Curves as in Fig.~3.}
\label{fig_shapdis_C}
\end{center}\end{figure}

The numerical results, Fig.~\ref{fig_shapdis_C}, show that the
estimated power correction is very large in this case, dominating
over the perturbative prediction even at $Q=50$ GeV. The size and
uncertainty in the correction suggest that, like the jet broadening,
the $C$-parameter is not a good variable for determining $\as$. 

\mysection{Summary}
In this paper we have investigated several infrared-safe
variables which characterize the shapes of
DIS final states in the current hemisphere of the Breit
frame, where one avoids as far as possible complications
associated with the target remnant.  We have presented
numerical predictions to leading order in perturbation
theory, together with estimates of leading non-perturbative
power corrections, which are predicted to be proportional
to $1/Q$, modulo logarithmic $Q$-dependence.
The assumption of an
approximately universal low-energy effective coupling
allowed us to relate the magnitudes of the corrections
to those in $\ee$ annihilation. We found that they are
expected to be largest, and most uncertain, for the
current jet broadening and $C$-parameter, and so these
observables are probably not suitable for determination
of the perturbative strong coupling $\as$. The current
jet thrust and mass should have power corrections that
are smaller and under better control. When higher-order
predictions for these quantities are available, our
predictions of the power corrections can also be
refined, and it should be possible to measure both
$\as$ and the relevant non-perturbative parameter,
$\a0(\mI)$ in Eq.~\re{A1exp}, from these quantities.

\section*{Acknowledgements}
M.D.\ acknowledges the financial support of Trinity College, Cambridge.
We thank S.\ Catani, Yu.L.\ Dokshitzer, G.\ Marchesini, H.-U.\ Martyn
and E.\ Mirkes for helpful conversations.

\par \vskip 1ex
\noindent{\large\bf References}
\begin{enumerate}
\item\label{jetexp}
H1 Collaboration, T.\ Ahmed et. al., \pl{346}{415}{95};\\
ZEUS Collaboration, M.\ Derrick et al., \pl{363}{201}{95}.
\item\label{jetalg}
J.G.\ K\"orner, E.\ Mirkes and G.\ Schuler, \ijmp{A4}{1781}{89};\\
T.\ Brodkorb, J.G.\ K\"orner, E.\ Mirkes, and G.\ Schuler,
\zp{44}{415}{89};\\
T.\ Brodkorb and J.G.\ K\"orner, \zp{54}{519}{92};\\
T.\ Brodkorb and E.\ Mirkes, \zp{66}{141}{95};\\
D.\ Graudenz \pl{256}{518}{92}, \pr{49}{3291}{94}.

\item\label{ktalg}
S.\ Catani, Y.L.\ Dokshitzer and B.R.\ Webber, \pl{285}{291}{92}.          

\item\label{MZ}
E.\ Mirkes and D.\ Zeppenfeld, \app{B27}{1393}{96},
\npps{51C}{273}{96}, TTP-96-30 [hep-ph/9608201],
MADPH-96-961 [hep-ph/9609274].

\item\label{CatSey}
S.\ Catani and M.H.\ Seymour, \np{485}{291}{97},
CERN-TH/96-239 [hep-ph/9609237],
CERN-TH/96-240 [hep-ph/9609521].

\item\label{hadro}
       B.R.\ Webber, \pl{339}{148}{94}.
\item\label{DokWeb}
       Yu.L.\ Dokshitzer and B.R.\ Webber, \ib{352B}{451}{95}.
\item\label{BPY}
       Yu.L.\ Dokshitzer, G.\ Marchesini and B.R.\ Webber,
       \np{469}{93}{96}.
\item\label{NaSey}
       P.\ Nason and M.H.\ Seymour, \np{454}{291}{95}.
\item\label{KorSte}
  G.P.\ Korchemsky and G.\ Sterman, in {\em Proc.\ 30th Rencontres de Moriond,
  Meribel-les-Allues, France, 1995} [hep-ph/9505391];\\
    R.\ Akhoury and V.I.\ Zakharov, \pl{357}{646}{95}, \np{465}{295}{96};
     V.M.\ Braun, NORDITA-96-65P [hep-ph/9610212];
     M.\ Beneke,  SLAC-PUB-7277  [hep-ph/9609215].
\item\label{renormalons}
       For reviews and classic references see
       V.I.\ Zakharov, \np{385}{452}{92} and 
       A.H.\ Mueller, in {\em QCD 20 Years Later}, vol.~1
       (World Scientific, Singapore, 1993).
\item\label{alfaeff} 
     G.\ Grunberg, \pl{372}{121}{96}, CPTH-PC463-0896 [hep-ph/9608375];\\ 
     D.V.\ Shirkov and I.L.\ Solovtsov, Dubna preprint, April 1996
     [hep-ph/9604363]. 
\item\label{shapexp}
  DELPHI Collaboration, P.\ Abreu et al., \zp{73}{229}{97}.
\item\label{fragren}
       M.\ Dasgupta and B.R.\ Webber, \np{484}{247}{97};\\
       M.\ Beneke, V.M.\ Braun and L.\ Magnea, SLAC-PUB-7274
       [hep-ph/9609266], CERN-TH/96-362 [hep-ph/9701309].
\item\label{DISren}
       E.\ Stein, M.\ Meyer-Hermann, L.\ Mankiewicz and A.\ Sch\"afer,
       \pl{376}{177}{96};
       M.\ Meyer-Hermann, M.\ Maul, L.\ Mankiewicz,
       E.\ Stein and A.\ Sch\"afer, \pl{383}{463}{96},
       \ib {393B}{487}{97} (E);   M.\ Maul, E.\ Stein, A.\ Sch\"afer
       and L.\ Mankiewicz, TUM-T39-96-29 [hep-ph/9612300];\\
       M.\ Dasgupta and B.R.\ Webber, \pl{382}{273}{96}.

\item\label{WParis}
  B.R.\ Webber in {\em Proc.\ Workshop on Deep Inelastic
  Scattering and QCD, Paris, 1995} [hep-ph/9510283].

\item\label{SWZ}
   K.H.\ Streng, T.F.\ Walsh and P.M.\ Zerwas, \zp{2}{237}{79}.
\item\label{PeRu}
   R.D.\ Peccei and R.\ R\"uckl, \np{162}{125}{80}.
\item\label{GDKT}
   L.V.\ Gribov, Yu.L.\ Dokshitzer, S.I.\ Troyan and V.A.\ Khoze,
   \spj{68}{1303}{88}.

\item\label{frag}
ZEUS Collaboration, M.\ Derrick et al., \zp{67}{93}{95};\\ 
H1 Collaboration, S.\ Aid et al., \np{445}{3}{95}. 

\item\label{CTW}
S.\ Catani, G.\ Turnock and B.R.\ Webber, \pl{295}{269}{92}.          

\item\label{ERT}
R.K.\ Ellis, D.A.\ Ross and A.E.\ Terrano, \np{178}{421}{81}.

\item\label{BBB}
   M.\ Beneke, V.M.\ Braun and V.I.\ Zakharov, \prl{73}{3058}{94};\\
   P.\ Ball, M.\ Beneke and V.M.\ Braun, \np{452}{563}{95};\\
   M.\ Beneke and V.M.\ Braun, \np{454}{253}{95}.
\item\label{SVZ}
    M.A.\ Shifman, A.I.\ Vainstein and V.I.\ Zakharov, 
    \np{147}{385,448,519}{79};\\
       {\it Vacuum Structure and QCD Sum Rules: Reprints},
       ed.\ M.A.\ Shifman (North-Holland, 1992:
       Current Physics, Sources and Comments, v.\ 10).
\item\label{CMW}
S.\ Catani, G.\ Marchesini and B.R.\ Webber, \np{349}{635}{91}.          
\item\label{MRSA}
       A.D. Martin, R.G. Roberts and W.J. Stirling, \pr{50}{6734}{94}.
\end{enumerate}
\end{document}